\documentstyle[11pt]{article}
\columnwidth\textwidth

\title{Perturbative gauge invariance: electroweak theory II}
\author{Andreas Aste, G\"unter Scharf\thanks{Work supported by Swiss National
Science Foundation} \\Institut f\"ur Theoretische Physik der Universit\"at
Z\"urich, \\ Winterthurerstr. 190, CH-8057 Z\"urich, Switzerland \and Michael
D\"utsch\thanks{Work supported by Alexander von Humboldt
Foundation}\\II.Institut f\"ur Theoretische Physik, Universit\"at Hamburg\\
Luruper Chaussee 149, D-22761 Hamburg, Germany} \date{}

\begin{document} \maketitle \begin{abstract}  A recent construction of the
electroweak theory, based on perturbative quantum gauge invariance 
alone, is extended to the
case of more generations of fermions with arbitrary mixing. The
conditions implied by second order gauge invariance lead to an isolated solution
for the fermionic couplings in agreement with the standard model.  Third order
gauge invariance determines the Higgs potential. The resulting massive gauge
theory is manifestly gauge invariant, after construction. \end{abstract}
\newpage 

\def\d{\partial}\def\=d{\,{\buildrel\rm def\over =}\,}
\def\dh{\mathop{\vphantom{\odot}\hbox{$\partial$}}} \def\dl{\dh^\leftrightarrow}
\def\sqr#1#2{{\vcenter{\vbox{\hrule height.#2pt\hbox{\vrule width.#2pt
height#1pt \kern#1pt \vrule width.#2pt}\hrule height.#2pt}}}}
\def\w{\mathchoice\sqr45\sqr45\sqr{2.1}3\sqr{1.5}3\,} \def\eps{\varepsilon}
\def\oe{\overline{\rm e}} \def\onu{\overline{\nu}}
\def\ds{\hbox{\rlap/$\partial$}} \def\psq{{\overline{\psi}}} \def\tr{{\rm tr}}

\section{Introduction} In a recent paper \cite{1} we have described a
formulation of the electroweak theory in terms of 
asymptotic fields alone. Since
in our causal approach the asymptotic gauge fields are massive from the very
beginning (except the photon), spontaneous symmetry breaking plays no role.

Let us summarize the strategy. The fundamental problem in gauge theories is 
the appearance of {\it unphysical fields} in the gauge potentials. An 
additional main difficulty is that already the quantization of the massive
\footnote{This problem appears also for the massless free gauge fields.
The peculiarity of the massive (free) theory is that we need the unphysical 
bosonic scalars $\Phi_a$ to restore the nilpotency of $Q$.} free gauge
fields in a $\lambda$-gauge requires an indefinite metric space. To
solve these two problems in the {\it free} theory one has to select the 
space of physical states ${\cal H}_{\rm phys}$ which must be a (pre) Hilbert
space, i.e. the inner product must be positive definite. There are two
popular methods at hand: the Gupta Bleuler method (which works well for 
abelian gauge theories only) and the BRST-formalism \cite{15}. We take the 
latter in the form of Kugo and Ojima \cite{16}: the physical (pre) Hilbert
space is defined as the cohomology of a nilpotent operator $Q$, 
i.e. ${\cal H}_{\rm phys}={{\rm Ker}\,Q\over {\rm Ran}\,Q}$. To define such 
an operator we introduce unphysical fields: each gauge field $A_a^\mu$ gets 
three scalar partners, the fermionic fields $u_a,\,\tilde u_a$ 
("ghost fields") and the bosonic $\Phi_a$. Then
$$Q\=d \int d^3x\,(\d_{\nu}A_a^{\nu}+m_a\Phi_a){\dl}_0u_a,$$
does the job.

To overcome the presence of unphysical fields in the {\it interacting theory}
one has additionally to prove that they {\it decouple}. The latter means 
that in the adiabatic limit (if it exists) the $S$-matrix induces a 
well-defined unitary operator. For this purpose one needs a notion of 
gauge invariance. The most natural formulation which yields the decoupling
is to require that $Q$ commutes with the $S$-matrix in the adiabatic limit
(if it exists). This is essentially the content of our formulation of
gauge invariance
$$d_QT_n\=d [Q,T_n]=i\sum_{l=1}^n{\d\over\d x_l^\nu}T^\nu_{n/l} 
(x_1,\ldots x_l\ldots x_n)\eqno(1.0)$$
(where we use the notations of \cite{1}), which is independent of the 
adiabatic limit, i.e. it makes sense also in theories in which this limit 
does not exist. This is a pure quantum formulation of gauge invariance.
It has turned out that this symmetry requirement (1.0) cannot be satisfied
with the fields at hand so far. By introducing an additional scalar field
(the "Higgs field"), which is physical, gauge invariance can be saved.
In addition (1.0) is sufficiently strong to determine the couplings of all
fields.

The main steps in this construction are the following. Starting from the pure
Yang - Mills coupling $T_1^A$ at first order, gauge invariance requires 
couplings $T_1^u$ of the 
ghost fields $u_a\,\tilde u_a$ and fixes them. This step
is the same as in the massless theory \cite{2}. However, the massive gauge
fields need couplings to the bosonic scalar partners $\Phi_a$ 
to compensate mass terms in the first order
gauge variation. Gauge invariance then determines the couplings $T_1^\Phi$ of
these unphysical scalars to the gauge fields and to the ghosts. But the
resulting theory would not be gauge invariant at second order. This requires 
the "Higgs field" $\Phi_0$, its coupling $T_1^{\Phi_0}$
to all other fields is again determined by gauge invariance. The coupling to
leptons is uniquely fixed by first and second order gauge invariance if one
family of fermions is considered \cite{1}. However, it is well known that gauge
invariance is then violated at third order by the triangular anomalies. Hence,
the theory must still be enlarged by including at least one complete generation
of leptons and quarks. In what follows we consider the general situation of
arbitrarily many generations of fermions with arbitrary mixing. 
It is our aim to
analyse what gauge invariance has to say about the couplings in this case.

Assuming that the reader is familiar with the main results of \cite{1}, we
present in the following section the general ansatz for the fermionic coupling
$T_1^F$ and determine the constraints imposed by gauge invariance at first
order. At second order (Sect.3) we obtain a long list of interesting conditions
for the coupling matrices. At present we cannot control all solutions of this
system. But inserting a general ansatz which is valid in the vicinity of the
standard model, we get a unique solution that agrees with the standard model. In
Sect.4 we briefly discuss gauge invariance of the third order triangular graphs.
We recover the well-known fact that the cancellation of the axial anomalies
restricts the charges of the fermions.

In Sect.5 we show how gauge invariance of third order tree graphs determines the
last free parameters in the scalar coupling. In this way we {\it derive} the
Goldstone - Higgs potential. Other approaches where this
quartic double-well potential is not postulated but deduced, are based 
on methods of non-commutative geometry which have been developed by 
Connes \cite{3} and
others \cite{4,5,6,7,12}, or on superconnections \cite{13}. 
The scalar potential comes out in the form
$$V(\Phi_0)=\lambda (\Phi_0^4-4a\Phi_0^3+4a^2\Phi_0^2),\eqno(1.1)$$ with
$$a={2\over g}m_W,\eqno(1.2)$$ where $m_W$ is the $W$-mass and $g$ the universal
gauge coupling constant.  The physical 'Higgs' field $\Phi_0$ is realized in an
ordinary Fock representation with unique vacuum and vacuum expectation value
$\langle \Phi_0\rangle=0$. 
(In the usual convention $-\Phi_0$ is used instead of
$\Phi_0$.) If we introduce the shifted field 
$$\varphi=\Phi_0-a,\eqno(1.3)$$ 
the potential assumes the usual symmetric form \cite{8}
$$V=\lambda(\varphi^2-a^2)^2.\eqno(1.4)$$ 
Now the vacuum expectation value
$\langle\varphi\rangle = -a$ is different from 0. If the shift (1.3) is carried
out everywhere in the scalar couplings (see (3.20) of \cite{1}) we get 
quadratic
mass terms (with the wrong sign) for the already massive gauge bosons. These
quadratic terms can simply be resummed which changes the W- and Z-masses from
their finite values into zero! That means, we have inverted the 
Higgs mechanism. This gives the connection of our approach with the
standard theory. However, this connection does not enter our construction.

\section{Fermionic coupling and first order gauge invariance}

We start from the following generalization of the simple leptonic electroweak
coupling to more than one family
$$T_1^F=ig\sum_{j,k}\Bigl\{W_\mu^+b^1_{jk}\oe_j\gamma^\mu\nu_k+W_\mu^+
b^{\prime 1}_{jk}\oe_j\gamma^\mu\gamma^5\nu_k\eqno(2.1.1)$$
$$+W_\mu^-b^2_{jk}\onu_j\gamma^\mu{\rm e}_k+W_\mu^-b^{\prime 2}_{jk}
\onu_j\gamma^\mu\gamma^5{\rm e}_k\eqno(2.1.2)$$ $$+Z_\mu
b^3_{jk}\oe_j\gamma^\mu{\rm e}_k+Z_\mu b^{\prime 3}_{jk}\oe_j
\gamma^\mu\gamma^5{\rm e}_k\eqno(2.1.3)$$ $$+Z_\mu
b^4_{jk}\onu_j\gamma^\mu\nu_k+Z_\mu b^{\prime 4}_{jk}\onu_j
\gamma^\mu\gamma^5\nu_k\eqno(2.1.4)$$ $$+A_\mu b^5_{jk}\oe_j\gamma^\mu{\rm
e}_k+A_\mu b^{\prime 5}_{jk}\oe_j \gamma^\mu\gamma^5{\rm e}_k\eqno(2.1.5)$$
$$+A_\mu b^6_{jk}\onu_j\gamma^\mu\nu_k+A_\mu b^{\prime 6}_{jk}\onu_j
\gamma^\mu\gamma^5\nu_k\eqno(2.1.6)$$
$$+\Phi^+c^1_{jk}\oe_j\nu_k+\Phi^+c^{\prime 1}_{jk}\oe_j\gamma^5\nu_k
+\Phi^-c^2_{jk}\onu_j{\rm e}_k+\Phi^-c^{\prime 2}_{jk}\onu_j\gamma^5 {\rm
e}_k\eqno(2.1.7)$$ $$+\Phi_3c^3_{jk}\oe_j{\rm e}_k+\Phi_3c^{\prime
3}_{jk}\oe_j\gamma^5 {\rm e}_k+\Phi_3c^4_{jk}\onu_j\nu_k+\Phi_3c^{\prime
4}_{jk}\onu_j \gamma^5\nu_k\eqno(2.1.8)$$ $$+\Phi_0c^0_{jk}\oe_j{\rm
e}_k+\Phi_0c^{\prime 0}_{jk}\oe_j\gamma^5 {\rm
e}_k+\Phi_0c^5_{jk}\onu_j\nu_k+\Phi_0c^{\prime 5}_{jk}\onu_j
\gamma^5\nu_k\Bigl\}.\eqno(2.1.9)$$ 
Here we have used the same notation as in
\cite{1} (5.1): All products of field operators throughout are normally
ordered (Wick monomials). $W, Z, A$ denote the gauge fields and 
$\Phi^\pm=(\Phi_1\pm i\Phi_2)/\sqrt{2}$, $\Phi_3$ the unphysical scalars.
$\Phi_0$ is the physical
scalar, but ${\rm e}_j (x)$ stands for the electron-, muon-, tau-fields, as well
as for the quark fields d, s, b, and $\nu_k(x)$ represents the corresponding
neutrini and the other quark fields u, c, t. In \cite{1} we have only considered
the coupling to leptons. There the terms (2.1.6) are missing because the
neutrini have vanishing electric charge, but here, for the quark couplings, we
must include them. We also assume that the asymptotic Fermi fields
fulfil the Dirac equations $$\ds{\rm e_j}=-im_j^e{\rm e_j},\quad
\d_\mu(\oe_j\gamma^\mu)=im_j^e\oe_j$$ $$\ds\nu_k=-im_k^\nu\nu_k,\quad
\d_\mu(\onu_k\gamma^\mu)=im_k^\nu\onu_k, \eqno(2.2)$$ 
with arbitrary non-vanishing unequal masses $m^e_j\ne m^e_k\ne 0$, 
$m^\nu_j\ne m^\nu_k\ne 0$ for all $j\ne k$.
We do not use further information about the multiplett
structure of the fermions.

According to \cite{1} the gauge structure is introduced as follows. We define a
gauge charge $$Q\=d \int
d^3x\sum_{a=0}^3(\d_{\nu}A_a^{\nu}+m_a\Phi_a){\dl}_0u_a,$$ where $A_a^\nu$
stands for $A^\nu, W_1^\nu, W_2^\nu, Z^\nu$ and $u_a$ for the corresponding
ghosts, and a gauge variation $$d_Q F\=d QF-(-1)^{n_F}FQ.$$ Then the gauge
variations of the asymptotic gauge fields are given by $$d_QA^\mu =i\d^\mu
u_0,\> d_Q W_{1,2}^\mu=i\d^\mu u_{1,2},\> d_Q Z^\mu=i\d^\mu u_3,$$ and for the
Higgs and unphysical scalar fields
$$d_Q\Phi_0=0,\>d_Q\Phi_{1,2}=im_Wu_{1,2},\>d_Q\Phi_3=im_Zu_3$$ and finally for
the fermionic ghosts $$d_Qu_a=0,\quad a=0,1,2,3$$ $$d_Q\tilde u_0=-i\d_\mu
A^\mu,\>d_Q\tilde u_{1,2}=-i(\d_\mu W_{1,2} ^\mu+m_W\Phi_{1,2})$$ $$d_Q\tilde
u_3=-i(\d_\mu Z^\mu+m_Z\Phi_3).\eqno(2.3)$$ The gauge variations of the
fermionic matter fields vanish.

First order gauge invariance means that the gauge variation of
$$T_1=T_1^A+T_1^u+T_1^\Phi+T_1^{\Phi_0}+T_1^F$$ has divergence form
$$d_QT_1(x)=i\d_\mu T_{1/1}^\mu.\eqno(2.4)$$ To verify this for the fermionic
coupling $T_1^F$ we calculate the gauge variation of (2.1) and take out the
derivatives of the ghost fields. In the additional terms with derivatives on the
matter fields we use the Dirac equations (2.2):
$$d_QT_1^F=-g\sum_{j,k}\d_\mu\Bigl\{u^+(b^1_{jk}\oe_j\gamma^\mu\nu_k+ b^{\prime
1}_{jk}\oe_j\gamma^\mu\gamma^5\nu_k) +u^-(b^2_{jk}\onu_j\gamma^\mu{\rm
e}_k+b^{\prime 2}_{jk} \onu_j\gamma^\mu\gamma^5{\rm e}_k)\eqno(a)$$
$$+u_3(b^3_{jk}\oe_j\gamma^\mu{\rm e}_k+b^{\prime 3}_{jk}\oe_j
\gamma^\mu\gamma^5{\rm e}_k +b^4_{jk}\onu_j\gamma^\mu\nu_k+b^{\prime
4}_{jk}\onu_j \gamma^\mu\gamma^5\nu_k)\eqno(b)$$
$$+u_0(b^5_{jk}\oe_j\gamma^\mu{\rm e}_k+b^{\prime 5}_{jk}\oe_j
\gamma^\mu\gamma^5{\rm e}_k +b^6_{jk}\onu_j\gamma^\mu\nu_k+b^{\prime
6}_{jk}\onu_j \gamma^\mu\gamma^5\nu_k)\Bigl\}\eqno(c)$$
$$+g\Bigl\{iu^+[b^1_{jk}(m_j^e-m_k^\nu)\oe_j\nu_k+ b^{\prime
1}_{jk}(m_j^e+m_k^\nu)\oe_j\gamma^5\nu_k]\eqno(d)$$
$$+iu^-[b^2_{jk}(m_j^\nu-m_k^e)\onu_j{\rm e}_k+b^{\prime 2}_{jk}
(m_j^\nu+m_k^e)\onu_j\gamma^5{\rm e}_k]$$ $$+iu_3[b^3_{jk}(m_j^e-m_k^e)\oe_j{\rm
e}_k+b^{\prime 3}_{jk}(m_j^e+m_k^e) \oe_j\gamma^5{\rm e}_k$$
$$+b^4_{jk}(m_j^\nu-m_k^\nu)\onu_j\nu_k+b^{\prime 4}_{jk}(m_j^\nu+m_k^\nu)
\onu_j\gamma^5\nu_k]$$ $$+iu_0[b^5_{jk}(m_j^e-m_k^e)\oe_j{\rm e}_k+b^{\prime
5}_{jk}(m_j^e+ m_k^e)\oe_j\gamma^5{\rm e}_k$$
$$+b^6_{jk}(m_j^\nu-m_k^\nu)\onu_j\nu_k+b^{\prime 6}_{jk}(m_j^\nu+m_k^\nu)
\onu_j\gamma^5\nu_k]\Bigl\}$$ $$+m_W[u^+c^1_{jk}\oe_j\nu_k+u^+c^{\prime
1}_{jk}\oe_j\gamma^5\nu_k +u^-c^2_{jk}\onu_j{\rm e}_k+u^-c^{\prime
2}_{jk}\onu_j\gamma^5 {\rm e}_k]$$ $$+m_Zu_3[c^3_{jk}\oe_j{\rm e}_k+c^{\prime
3}_{jk}\oe_j\gamma^5 {\rm e}_k+c^4_{jk}\onu_j\nu_k+c^{\prime 4}_{jk}\onu_j
\gamma^5\nu_k]\Bigl\}.\eqno(2.5)$$

Now, to have first order gauge invariance, the terms ($d$) until the end of
(2.5) which are not of divergence form must cancel. This implies
$$b_{jk}^{\prime 5}=0=b_{jk}^{\prime 6},\>\forall j,k,\quad
b_{jk}^5=0=b_{jk}^6\>{\rm for}\>j\ne k\eqno(2.6)$$ and $$c_{jk}^1={i\over
m_W}(m_j^e-m_k^\nu)b_{jk}^1,\quad c_{jk}^{\prime 1}={i\over
m_W}(m_j^e+m_k^\nu)b_{jk}^{\prime 1}$$ $$c_{jk}^2={i\over
m_W}(m_j^\nu-m_k^e)b_{jk}^2,\quad c_{jk}^{\prime 2}={i\over
m_W}(m_j^\nu+m_k^e)b_{jk}^{\prime 2}$$ $$c_{jk}^3={i\over
m_Z}(m_j^e-m_k^e)b_{jk}^3,\quad c_{jk}^{\prime 3}={i\over
m_Z}(m_j^e+m_k^e)b_{jk}^{\prime 3}$$ $$c_{jk}^4={i\over
m_Z}(m_j^\nu-m_k^\nu)b_{jk}^4,\quad c_{jk}^{\prime 4}={i\over
m_Z}(m_j^\nu+m_k^\nu)b_{jk}^{\prime 4}.\eqno(2.7)$$ The result (2.6) means that
the photon has no axial-vector coupling and no mixing in the vector coupling.
This is due to the fact that it has no scalar partner because it is massless.

\section{Gauge Invariance at Second Order}

As discussed in Sect.4 and 5 of \cite{1}, the essential problem in 
second order gauge
invariance is whether the anomalies in the tree graphs cancel out.  These
anomalies are the local terms in $\d_\nu^x T_{2/1}^\nu
\vert^0_{\rm tree}(x,y)+\d_\nu^y T_{2/2}^\nu\vert^0_{\rm tree} (x,y)$
and come from two sources. First, if the terms ($a$)-($c$) in (2.5) are
combined with the terms in (2.1) by a fermionic contraction we get the Feynman
propagator $S^F_m(x-y)$ in $T_{2/1}^\nu
\vert^0_{\rm tree}(x,y)$ and $T_{2/2}^\nu\vert^0_{\rm tree} (x,y)$. Taking
the divergence with respect to the $Q$-vertex a $\delta$-distribution is
generated due to
$$i\d_\mu^x\gamma^\mu S_m^F(x-y)=mS_m^F(x-y)+\delta(x-y).  \eqno(3.1)$$ 
This $\delta$-term is
the anomaly. Secondly, we can perform a bosonic contraction between the 
terms in
(2.1) and the terms (4.5) in \cite{1} which are the anomaly-producing part in
$T_{1/1}^{A\Phi\,\mu}$, coming from the Yang-Mills and scalar couplings:
$$T_{1/1}^{A\Phi\,\mu}\vert_{\rm an}=ig\Bigl\{\sin\Theta(u_1W_2^\nu-
u_2W_1^\nu) \d^\mu A_\nu \eqno(3.2.1)$$ $$+\sin\Theta
(u_2A^\nu-u_0W_2^\nu)\d^\mu W_{1\nu}+\sin\Theta (u_0 W_1^\nu-u_1A^\nu)\d^\mu
W_{2\nu} \eqno(3.2.2)$$ $$+\cos\Theta\Bigl[(u_2Z^\nu-u_3W_2^\nu)\d^\mu W_{1\nu}
+(u_3W_1^\nu-u_1Z^\nu)\d^\mu W_{2\nu}\eqno(3.2.3)$$
$$+(u_1W_2^\nu-u_2W_1^\nu)\d^\mu Z_\nu\Bigl]\eqno(3.2.4)$$ $$+\sin\Theta
(u_0u_1\d^\mu\tilde u_2+u_2u_0\d^\mu\tilde u_1+ u_1u_2\d^\mu\tilde
u_0)\eqno(3.2.5)$$ $$+\cos\Theta (u_2u_3\d^\mu\tilde u_1+u_3u_1\d^\mu\tilde u_2
+u_1u_2\d^\mu\tilde u_3)\eqno(3.2.6)$$ $$+\sin\Theta
u_0(\Phi_2\d^\mu\Phi_1-\Phi_1\d^\mu\Phi_2)+\Bigl(1-{m_Z^2 \over
2m_W^2}\Bigl)\cos\Theta u_3(\Phi_2\d^\mu\Phi_1-\Phi_1\d^\mu\Phi_2)\eqno(3.2.7)$$
$$+{1\over 2}{m_Z\over m_W}\cos\Theta\Bigl[(u_2\Phi_1-u_1\Phi_2)\d^\mu\Phi_3
+u_1\Phi_3\d^\mu\Phi_2-u_2\Phi_3\d^\mu\Phi_1\Bigl] \eqno(3.2.8)$$ $$+{1\over
2}u_1(\Phi_0\d^\mu\Phi_1-\Phi_1\d^\mu\Phi_0) +{1\over
2}u_2(\Phi_0\d^\mu\Phi_2-\Phi_2\d^\mu\Phi_0)\eqno(3.2.9)$$ $$+{1\over
2\cos\Theta}u_3(\Phi_0\d^\mu\Phi_3- \Phi_3\d^\mu\Phi_0)\Bigl\}.\eqno(3.2.10)$$

To give a representative example, we calculate the anomalies with external 
field
operators $u_3\Phi_3\oe_j\gamma^5{\rm e}_k$. Combining the first term in (2.5)
($b$) with the second term in (2.1.8) we get an anomaly from the 
contraction of ${\rm e}_{k'}(x)$ with $\oe_{j'}(y)$ ($S^F[{\rm e}_{k'}(x),
\oe_{j'}(y)]$ denotes the corresponding Feynman propagator)
$$-iu_3(x)b^3_{jk'}\oe_j(x)\gamma^\mu
S^F[{\rm e}_{k'}(x),\oe_{j'}(y)]\gamma^5{\rm e}_k(y) \Phi_3(y)
c^{\prime 3}_{j'k}.$$ 
It results the anomaly $\sim -ib^3c^{\prime 3}\delta(x-y)$ involving
the matrix product of $b^3$ with $c^{\prime 3}$. Combining the two terms with
reversed order, we obtain $c^{\prime 3}b^3$ with a different sign so that both
terms together yield the commutator $i[c^{\prime 3},b^3]$. 
Similarly, the second
term in (2.5) ($b$) together with the first term in (2.1.8) gives the
anticommutator $\{c^3,b^{\prime 3}\}$, because the $\gamma^5$ is at a different
place. An anomaly of the second source comes from the last term in (3.2.10)
contracted by the two $\Phi_0$-fields with the second term in (2.1.9):
$$-{i\over 2\cos\Theta}u_3(x)\Phi_3(x)D^F[\d^\mu\Phi_0(x),\Phi_0(y)]
\oe_j(y)\gamma^5 {\rm e}_k(y)c_{jk}^{\prime 0}.$$ 
Altogether we obtain the following matrix equation
$$-{1\over 2\cos\Theta}c^{\prime 0}=i[c^{\prime 3},b^3]+i\{c^3, b^{\prime
3}\}.\eqno(3.3)$$

We now give the complete list of all second order conditions. We specify the
corresponding external legs, then the origin of the terms is pretty clear. Every
combination of external field operators has a corresponding one with an
additional $\gamma^5$. To save space we do not write down the external legs once
more for the $\gamma^5$-term.

$u_0\Phi_0\oe{\rm e}:\>[c^0,b^5]=0,\quad [c^{\prime 0},b^5]=0$

$u_0\Phi_3\oe{\rm e}:\>[c^3,b^5]=0,\quad [c^{\prime 3},b^5]=0$

$u_0Z\oe\gamma{\rm e}:\>[b^3,b^5]=0,\quad [b^{\prime 3},b^5]=0$

$u_0\Phi_0\onu\nu:\>[c^5,b^6]=0,\quad [c^{\prime 5},b^6]=0$

$u_0\Phi_3\onu\nu:\>[c^4,b^6]=0,\quad [c^{\prime 4},b^6]=0$

$u_0Z\onu\gamma\nu:\>[b^4,b^6]=0,\quad [b^{\prime 4},b^6]=0$ \qquad (3.4)

$u^+A\oe\gamma\nu:\>\sin\Theta b^1=b^5b^1-b^1b^6,\quad \sin\Theta b^{\prime
1}=b^5b^{\prime 1}-b^{\prime 1}b^6$

$u^-A\onu\gamma{\rm e}:\>\sin\Theta b^2=b^2b^5-b^6b^2,\quad \sin\Theta
b^{\prime 2}=b^{\prime 2}b^5-b^6b^{\prime 2}$ \qquad (3.5)

$u_0\Phi^+\oe\nu:\>\sin\Theta c^1=b^5c^1-c^1b^6,\quad \sin\Theta c^{\prime
1}=b^5c^{\prime 1}-c^{\prime 1}b^6$

$u_0\Phi^-\onu{\rm e}:\>\sin\Theta c^2=c^2b^5-b^6c^2,\quad \sin\Theta c^{\prime
2}=c^{\prime 2}b^5-b^6c^{\prime 2}$ \qquad (3.6)

$u_3W^+\oe\gamma\nu:\>\cos\Theta b^1=b^3b^1+b^{\prime 3}b^{\prime 1}
-b^1b^4-b^{\prime 1}b^{\prime 4},\quad \cos\Theta b^{\prime 1}=b^{\prime
3}b^1+b^3b^{\prime 1} -b^{\prime 1}b^4-b^1b^{\prime 4}$

$u_3W^-\onu\gamma{\rm e}:\>\cos\Theta b^2=b^2b^3+b^{\prime 2}b^{\prime 3}
-b^4b^2-b^{\prime 4}b^{\prime 2},\quad \cos\Theta b^{\prime 2}=b^{\prime
2}b^3+b^2b^{\prime 3} -b^{\prime 4}b^2-b^4b^{\prime 2}$

\rightline{(3.7)}

$u^+W^-\oe\gamma{\rm e}:\>\cos\Theta b^3=b^1b^2+b^{\prime 1}b^{\prime 2}
-\sin\Theta b^5,\quad \cos\Theta b^{\prime 3}=b^{\prime 1}b^2+b^1b^{\prime 2}$

$u^-W^+\onu\gamma\nu:\>\cos\Theta b^4=-b^2b^1-b^{\prime 2}b^{\prime 1}
-\sin\Theta b^6,\quad \cos\Theta b^{\prime 4}=-b^{\prime 2}b^1-b^2b^{\prime 1}
$\qquad (3.8)

$u^+\Phi_3\oe\nu:\>c^1=2(b^{\prime 1}c^{\prime 4}+c^{\prime 3}b^{\prime 1}
+c^3b^1-b^1c^4),\quad c^{\prime 1}=2(b^{\prime 1}c^4+c^3b^{\prime 1} +c^{\prime
3}b^1-b^1c^{\prime 4})$

$u^-\Phi_3\onu{\rm e}:\>c^2=2(-b^{\prime 2}c^{\prime 3}-c^{\prime 4} b^{\prime
2}-c^4b^2+b^2c^3),\quad c^{\prime 2}=2(-b^{\prime 2}c^3-c^4b^{\prime 2}
-c^{\prime 4}b^2+b^2c^{\prime 3})$

\rightline{(3.9)}

$u^+\Phi_0\oe\nu:\>\delta_1c^1=i(c^0b^1+c^{\prime 0}b^{\prime 1} +b^{\prime
1}c^{\prime 5}-b^1c^5),\quad \delta_1c^{\prime 1}=i(c^{\prime 0}b^1+c^0b^{\prime
1} +b^{\prime 1}c^5-b^1c^{\prime 5})$

$u^-\Phi_0\onu{\rm e}:\>\delta_1c^2=i(c^5b^2+c^{\prime 5} b^{\prime
2}+b^{\prime 2}c^{\prime 0}-b^2c^0),\quad \delta_1c^{\prime 2}=i(c^{\prime
5}b^2+c^5b^{\prime 2} +b^{\prime 2}c^0-b^2c^{\prime 0})$

\rightline{(3.10)}

$u_3\Phi_3\oe{\rm e}:\>\delta_3c^0=i[b^3,c^3]-i\{b^{\prime 3},c^{\prime
3}\},\quad\delta_3c^{\prime 0}=i[b^3,c^{\prime 3}]-i\{b^{\prime 3},c^3\}$

$u_3\Phi_0\oe{\rm e}:\>\delta_3c^3=i[c^0,b^3]+i\{c^{\prime 0},b^{\prime
3}\},\quad\delta_3c^{\prime 3}=i[c^{\prime 0},b^3]+i\{c^0 ,b^{\prime 3}\}$
\qquad (3.11)

$u_3\Phi_3\onu\nu:\>\delta_3c^5=i[b^4,c^4]-i\{b^{\prime 4},c^{\prime
4}\},\quad\delta_3c^{\prime 5}=i[b^4,c^{\prime 4}]-i\{b^{\prime 4},c^4\}$

$u_3\Phi_0\onu\nu:\>\delta_3c^4=i[c^5,b^4]+i\{c^{\prime 5},b^{\prime
4}\},\quad\delta_3c^{\prime 4}=i[c^{\prime 5},b^4]+i\{c^ 5,b^{\prime 4}\}$
\qquad (3.12)

$u^+\Phi^-\oe{\rm e}:\>\delta_1c^0=i(b^1c^2-b^{\prime 1}c^{\prime
2})-ic^3/2,\quad\delta_1c^{\prime 0}=i(b^1c^{\prime 2}-b^{\prime 1}c^2)
-ic^{\prime 3}/2$

$u^-\Phi^+\oe{\rm e}:\>\delta_1c^0=-i(c^1b^2+c^{\prime 1}b^{\prime
2})+ic^3/2,\quad\delta_1c^{\prime 0}=-i(c^{\prime 1}b^2+c^1b^{\prime 2})
+ic^{\prime 3}/2$\qquad (3.13)

$u^+\Phi^-\onu\nu:\>\delta_1c^5=i(b^2c^1-b^{\prime 2}c^{\prime
1})-ic^4/2,\quad\delta_1c^{\prime 5}=i(b^2c^{\prime 1}-b^{\prime 2}c^1)
-ic^{\prime 4}/2$

$u^-\Phi^+\onu\nu:\>\delta_1c^5=-i(c^2b^1+c^{\prime 2}b^{\prime
1})+ic^4/2,\quad\delta_1c^{\prime 5}=-i(c^{\prime 2}b^1+c^2b^{\prime 1})
+ic^{\prime 4}/2$\qquad (3.14)

$u_3\Phi^+\oe\nu:\>\delta_4c^1=b^3c^1-b^{\prime 3}c^{\prime 1}-c^1b^4-c^{\prime
1}b^{\prime 4},\quad\delta_4c^{\prime 1}=-b^{\prime 3}c^1+b^3c^{\prime
1}-c^1b^{\prime 4}-c^{\prime 1}b^4$

$u_3\Phi^-\onu{\rm e}:\>\delta_4c^2=-b^4c^2+b^{\prime 4}c^{\prime
2}+c^2b^3+c^{\prime 2}b^{\prime 3},\quad\delta_4c^{\prime 2}=b^{\prime
4}c^2-b^4c^{\prime 2}+c^2b^{\prime 3}+c^{\prime 2}b^3$,

\rightline{(3.15)}

where $$\delta_1={1\over 2},\quad\delta_3={1\over 2\cos\Theta},
\quad\delta_4=\cos\Theta-{1\over 2\cos\Theta}.$$ The terms with these $\delta$'s
and with the electroweak mixing angle obviously come from (3.2). There are
further combinations of external field operators which have not been written
down, because they give no new condition.

In case of one family, assuming $b^6=0$ and taking pseudounitarity into account
(\cite{1} (5.21)), the corresponding system of scalar equations has a unique
solution, which agrees with the lepton coupling of the standard model. The
solution of the above matrix equations (3.4-15) is not so simple. We start from
the equations (3.5). If we write these equations with matrix elements, using the
fact that $b^5$ and $b^6$ are diagonal (2.6), we easily conclude that $b^5$ and
$b^6$ are actually multiples of the unit matrix $$b^5=\alpha{\bf 1},\quad
b^6=(\alpha-\sin\Theta){\bf 1}.\eqno(3.16)$$ 
Here $\alpha$ is a free parameter (the electric charge of the upper
quarks or leptons)
and we have assumed that the matrices $b^1, b^{\prime 1}, b^2, b^{\prime 2}$
are nontrivial and non-diagonal. This is not a serious limitation because we shall see that these matrices are essentially the unitary mixing matrices.
For the leptons we assume $b_6=0$ instead. This first consequence of gauge 
invariance is
the universality of the electromagnetic coupling: the members $e_k(x)$  and
$\nu_k(x)$ of different generations all couple in the same way to the photon,
with a constant charge difference $q_e-q_\nu=g\sin\Theta$, which is the
electronic charge.

Next we turn to the conditions (3.11). It is convenient to introduce the
diagonal mass matrices $$m^e={\rm diag}(m^e_j),\quad m^\nu={\rm
diag}(m^\nu_j),\> j=1,\ldots n_g,$$ where $n_g$ is the number of generations.
Then (2.7) can be written as follows $$c^1={i\over m_W}(m^eb^1-b^1m^\nu),$$
$$c^{\prime 1}={i\over m_W}(m^eb^{\prime 1}+b^{\prime 1}m^\nu),\eqno(3.17) $$
etc. Then the first two equations in (3.11) read $$c^{\prime
0}={1\over\delta_3m_Z}\Bigl(2b^{\prime 3}m^eb^3-2b^3m^e b^{\prime 3}$$
$$+m^eb^{\prime 3}b^3+m^eb^3b^{\prime 3}-b^3b^{\prime 3}m^e-b^{\prime 3}
b^3m^e\Bigl)$$ $$c^0={1\over\delta_3m_Z}\Bigl(2b^{\prime 3}m^eb^{\prime
3}-2b^3m^eb^3 +(b^3)^2m^e+m^e(b^3)^2+(b^{\prime 3})^2m^e+m^e(b^{\prime
3})^2\Bigl).  \eqno(3.18)$$ Substituting this into the last two equations of
(3.11), we arrive at the following coupled matrix equations for $b^3, b^{\prime
3}$: $$\delta_3^2(m^eb^{\prime 3}+b^{\prime 3}m^e)=3b^{\prime 3}m^e(b^3)^2
+3(b^3)^2m^eb^{\prime 3}+3(b^{\prime 3})^2m^eb^{\prime 3}+3b^{\prime 3}
m^e(b^{\prime 3})^2$$ $$-3b^3m^eb^{\prime 3}b^3-3b^3b^{\prime
3}m^eb^3-3b^{\prime 3}b^3m^eb^3 -3b^3m^eb^3b^{\prime 3}$$ $$+m^eb^{\prime
3}(b^3)^2+m^eb^3b^{\prime 3}b^3+m^e(b^{\prime 3})^3+ m^e(b^3)^2b^{\prime 3}$$
$$+(b^3)^2b^{\prime 3}m^e+b^3b^{\prime 3}b^3m^e+b^{\prime 3}(b^3)^2m^e
+(b^{\prime 3})^3m^e\eqno(3.19)$$ $$\delta_3^2(m^eb^3-b^3m^e)=3(b^{\prime
3})^2m^eb^3+3(b^3)^2m^eb^3+ 3b^{\prime 3}m^eb^{\prime 3}b^3+3b^{\prime
3}m^eb^3b^{\prime 3}$$ $$-3b^3m^e(b^3)^2-3b^3b^{\prime 3}m^eb^{\prime
3}-3b^{\prime 3}b^3m^e b^{\prime 3}-3b^3m^e(b^{\prime 3})^2$$
$$+m^e(b^3)^3-(b^3)^3m^e+m^e(b^{\prime 3})^2b^3-(b^{\prime 3})^2b^3m^e$$
$$+m^eb^3(b^{\prime 3})^2-b^3(b^{\prime 3})^2m^e+m^eb^{\prime 3}b^3 b^{\prime
3}-b^{\prime 3}b^3b^{\prime 3}m^e.\eqno(3.20)$$

The coupled cubic equations (3.19-20) have many solutions in general. To
determine the solutions in the neighbourhood of the standard model, we
substitute $$b^3=\delta_3(\beta{\bf 1}+x),\quad b^{\prime 3}=\delta_3(\beta'
{\bf 1}+y),\quad \beta,\beta'\in {\bf C},\eqno(3.21)$$ 
and assume the matrices $x, y$ to be small so that only
terms linear in $x$ and $y$ must be taken with in (3.19-20). Then the equations
collapse to the simple form $$2\beta'(1-4\beta^{\prime 2})m+(1-12\beta^{\prime
2})(my+ym)=0 \eqno(3.22)$$ $$(1-12\beta^{\prime 2})(mx-xm)=0.\eqno(3.23)$$ Now,
(3.22) yields a unique solution if $\beta'=O(1)$ is assumed
$$\beta'={\eps_2\over 2},\quad\eps_2=\pm 1,\quad y=0,\eqno(3.24)$$ 
where the
last result follows by writing the vanishing anticommutator $\{m^e,y\}$ with
matrix elements, using $m^e_j>0, \forall j$. Then (3.23) implies
$$(m^e_i-m^e_k)x_{ik}=0,\quad \hbox{\rm no sum over}\>i,k\eqno(3.25)$$ 
that
means $x$ and $b^3$ are diagonal, taking into account that the masses 
$m^e_j$ are not
degenerate. All matrices in (3.18) commute, thus $$c^{\prime 0}=0\eqno(3.26)$$
$$c^0={4m^e\over\delta_3m_Z}(b^{\prime 3})^2={\delta_3\over m_Z}m^e,
\eqno(3.27)$$ and (3.11) gives $$c_3=0,\quad c^{\prime 3}=i\eps_2{\delta_3\over
m_Z}m^e.\eqno(3.28)$$

The same reasoning can be carried through for (3.12) which leads to $$b^{\prime
4}=-\eps_2{\delta_3\over 2},\quad c^4=0,\eqno(3.29)$$ for the sign see below,
and $$c^{\prime 4}=-i\eps_2{\delta_3\over m_Z}m^\nu,\quad c^{\prime 5}
=0\eqno(3.30)$$ $$c^5={\delta_3\over m_Z}m^\nu.\eqno(3.31)$$ With this knowledge
we turn to (3.9). Substituting $c^1$ in the first equation by (3.17) we arrive
at $$m^e(b^1-\eps_2b^{\prime 1})=(b^1-\eps_2b^{\prime 1})m^\nu,$$ leading to
$$b^{\prime 1}=\eps_2b^1,\eqno(3.32)$$ assuming non-degenerate masses again. In
the same way the second equation in (3.9) yields $$b^{\prime
2}=\eps_2b^2.\eqno(3.33)$$ This is the chiral coupling of all fermion
generations. The sign in (3.29) follows from (3.8).

Finally, from the four conditions (3.8) it is easy to conclude $$b^2={1\over
8}(b^1)^{-1}\eqno(3.34)$$ $$b^3={1\over\cos\Theta}\Bigl({1\over
4}-\alpha\sin\Theta\Bigl) \eqno(3.35)$$ $$b^4={1\over\cos\Theta}\Bigl(-{1\over
4}-\alpha\sin\Theta+\sin^2 \Theta\Bigl).\eqno(3.36)$$ This means that $x$ in
(3.21) is actually zero, so that {\it there is no other solution in the
neighbourhood of the standard model}. But solutions "far away" are not excluded.
All values of the $b$'s and $c$'s agree with the standard model \cite{8} for an
arbitrary number $n_g$ of generations. It is not hard to check that with the
results so determined all other second order conditions of gauge invariance are
satisfied. To finish this discussion we notice that pseudo-unitarity implies
$$b^{1+}=b^2={1\over 8}(b^1)^{-1},$$ hence $$b^1={1\over 2\sqrt{2}}V,\quad
b^2={1\over 2\sqrt{2}}V^+,\eqno(3.37)$$ where $V$ is an arbitrary unitary
matrix. This is the CKM mixing matrix \cite{8,9} for the quark coupling. A
similar mixing is possible for the charged leptonic currents. The recently 
observed signals of neutrino oscillations show that this mixing probably 
occurs.

\section{Gauge invariance at third order: axial anomalies}

Adler, Bell and Jackiw \cite{10} discovered that there exists a possibility to
violate gauge invariance at third order in the triangular graphs. This holds also true in the causal approach to gauge theory (\cite{14} sect.5.3). 
The anomalous
graphs contain one axial-vector and two vector couplings (VVA) or three
axial-vector couplings (AAA) of the fermions and three external gauge fields. To
have a more compact notation, we collect all fermionic matter fields into a big
vector $\psi=({\rm e},\mu,\tau,\nu_e,\nu_\mu,\nu_\tau)$ or $=(d,s,b,u,c,t),$
respectively. The gauge fields $A,W^+,W^-,Z$ are denoted by $A^\mu_a$ with
$a=0,+,-,3$. Then the coupling between fermions and gauge fields in (2.1) can be
written as 
$$T_1^{FA}=ig\Bigl\{\psq\gamma_\mu M_a\psi
A_a^\mu+\psq\gamma_\mu\gamma^5 M'_a\psi A_a^\mu\Bigl\},\eqno(4.1)$$ 
where $M_a$
stands for the following matrices of matrices: $$M_+=\pmatrix{0&b^1\cr
0&0\cr},\quad M_-=\pmatrix{0&0\cr b^2&0\cr}\eqno (4.2)$$ $$M_3=\pmatrix{b^3&0\cr
0&b^4\cr},\quad M_0=\pmatrix{b^5&0\cr 0&b^6\cr}, \eqno(4.3)$$ and similarly for
the axial-vector couplings, denoted by a prime.

Each triangular graph gives rise to two diagrams which differ by a permutation
of two vertices. Therefore, we have to compute the traces
$$\tr(M_aM_bM_c)+\tr(M_aM_cM_b)=\tr(M_a\{M_b,M_c\}),$$ where one or three $M$'s
must be axial-vector couplings with a prime.  It is well-known that the
cancellation of the axial anomalies relies on the compensation of these traces
in the sum of the leptonic and hadronic contributions. In this way third order
gauge invariance gives a further restriction of the quark coupling. To work this
out in detail, we consider the following cases.

{\it A. Case $(a=0,b=+,c=-)_{VVA}$:}

$$\tr (M_0\{M'_+,M_-\})+\tr (M_0\{M_+,M'_-\})=$$ $$=\tr[b^5(b^1b^{\prime
2}+b^{\prime 1}b^2)+b^6(b^2b^{\prime 1}+ b^{\prime 2}b^1)]$$ Using the results
of the last section, this is equal to $$=n_g{\eps_2\over
4}(2\alpha-\sin\Theta),\eqno(4.4)$$ where $n_g$ is the number of generations,
i.e. the dimension of the matrices $b^k$, and $\alpha$ is the charge of the
fermions in (3.16).  For leptons we have $b^6=0$, because the neutrini have no
electric charge, so that $$\alpha_L=\sin\Theta.\eqno(4.5)$$ Consequently, to
compensate the triangular anomaly proportional to (4.4), one needs the
compensation between leptons and quarks.  We assume equal number of families in
the lepton and quark sectors. Then for three colors of quarks one must have
$$2\alpha_L-\sin\Theta+3(2\alpha_Q-\sin\Theta)=0,\eqno(4.6)$$ which implies
$$\alpha_Q={1\over 3}\sin\Theta\eqno(4.7)$$ by (4.5).

In most textbooks the electric charge $\alpha_Q$ (4.7) of the d-, s-, b-quarks is put in and then, by requiring cancellation of the anomalies, one concludes 
that the number of families in the lepton and quark sectors must be equal. We have simply reversed the argument. 

{\it B. Case $(0,3,3)_{VVA}$:}

In this case the trace is simply given by
$$\tr(M_0\{M'_3,M_3\})=\tr[b^5(b^{\prime 3}b^3+b^3b^{\prime 3})+ b^6(b^{\prime
4}b^4+b^4b^{\prime 4})]$$ $$=n_g\eps_2{\delta_3\over\cos\Theta}\Bigl({1\over
4}-\sin^2\Theta \Bigl)\Bigl(2\alpha-\sin\Theta\Bigl).\eqno(4.8)$$ Due to the
same factor $(2\alpha-\sin\Theta)$ as in (4.4) the mechanism of compensation is
the same.

{\it C. Case $(3,+,-)_{VVA}$:}

Here the trace is equal to
$$\tr(M'_3\{M_+,M_-\}+M_3\{M'_+,M_-\}+M_3\{M_+,M'_-\}=$$
$$=\tr[b'_3b_1b_2+b'_4b_2b_1+b^3(b^1b^{\prime 2}+b^{\prime 1}b^2)
+b^4(b^2b^{\prime 1}+b^{\prime 2}b^1)]$$ $$=n_g{\eps_2\sin\Theta\over
4\cos\Theta}(\sin\Theta-2\alpha),\eqno(4.9)$$ with the same consequences as
before.

{\it D. Case $(3,+,-)_{AAA}$:}

Now we have to compute the trace $$\tr(M'_3\{M'_+,M'_-\}=\tr[b^{\prime
3}b^{\prime 1}b^{\prime 2}+ b^{\prime 4}b^{\prime 2}b^{\prime 1}]=0.$$

{\it E. Case $(0,0,3)_{VVA}$:}

Here the relevant trace is equal to $$\tr(M_0^2M'_3)=\tr[(b^5)^2b^{\prime
3}+(b^6)^2b^{\prime 4}$$ $$=n_g\eps_2{\delta_3\over
2}\sin\Theta(2\alpha-\sin\Theta).$$ Due to the same factor as in case A, the
compensation  between leptons and quarks is the same. This is also true in the
next case:

{\it F. Case $(3,3,3)_{VVA}$:}

$$\tr(M_3^2M'_3)=\tr[(b^3)^2b^{\prime 3}+(b^4)^2b^{\prime 4}]$$
$$=n_g\eps_2{\delta_3\over \cos\Theta}\Bigl(\sin^3\Theta-{\sin\Theta \over
2}\Bigl)(2\alpha-\sin\Theta).$$

{\it G. Case $(3,3,3)_{AAA}$:}

This final case is trivial: $$\tr(M'_3)^3=\tr[(b^{\prime 3})^3+(b^{\prime
4})^3]=0.$$ Summing up the axial anomalies completely cancel in each generation,
if and only if $\alpha_Q$ has the value (4.7).

\section{Gauge invariance at third order: tree graphs}

Third order tree graphs are not covered by the general inductive proof of gauge
invariance in massless Yang-Mills theories \cite{11}. They are part of the
beginning of the induction and need an explicit verification of gauge
invariance. The latter can easily be done in the massless Yang-Mills theory, by
using the fact that all couplings are of Yang-Mills type, that means
proportional to $f_{abc}$. But this is no longer true for the scalar couplings
in massive Yang-Mills theories \cite{1}. Therefore, it is not surprising that
gauge invariance of third order tree diagrams fixes the last free parameters in
the scalar coupling of the electroweak theory, namely the Higgs self-coupling
(Higgs potential).

Let us first discuss a simple special case of the standard model, the $U(1)$
Higgs model, which contains all essential features of the scalar self-coupling.
We consider one massive gauge field $W^\mu$ only, in interaction with one
unphysical ($\Phi$) and one physical scalar field $\Phi_0$. Therefore, we let
$W_1=W, u_1=u, \tilde u_1=\tilde u, \Phi_1=\Phi, m_W=m$ in (3.20) of \cite{1}
and omit the other fields. From (3.20.9, 10, 12) we then get $$T_1=i{g\over
2}\Bigl\{W^\nu(\Phi_0\d_\nu\Phi-\Phi\d_\nu\Phi_0) -mW_\nu W^\nu\Phi_0$$
$$+{m_H^2\over 2m}\Phi_0\Phi^2+m\tilde uu\Phi_0+2b\Phi_0^3\Bigl\}, \eqno(5.1)$$
and from (4.5.10) of \cite{1} or (3.2.9) $$T^\mu_{1/1}\vert_{\rm an}=-i{g\over
2}u\{\Phi\d^\mu\Phi_0-\Phi_0 \d^\mu\Phi\}\=d D_1+D_2.\eqno(5.2)$$

The calculation and compensation of the anomalies at second order can now be
directly taken over from the appendix in \cite{1}. Only the following two
sectors 1) and 7) appear:

\noindent {\it 1) Sector $(\Phi_0,\Phi_0,1,1)$}

Here we have found the two normalization terms (A.1, 2) $$N_1={i\over
4}g^2W_\nu W^\nu\Phi_0^2\delta(x-y)\eqno(5.3)$$ $$N_2=ig^2\Bigl({m_H^2\over
4m^2}-{3b\over 2m}\Bigl)\Phi_0^2 \Phi^2\delta(x-y).\eqno(5.4)$$

\noindent {\it 7) Sector $(1,1,1,1)$}

$$N_{10}={i\over 4}g^2W_\nu W^\nu\Phi^2\delta(x-y)\eqno(5.5)$$
$$N_{11}=-ig^2{m_H^2\over 16m^2}\Phi^4\delta(x-y)\eqno(5.6)$$ These results
(5.3-6) are valid for the $U(1)$ Higgs model, too, because all additional
couplings in the standard model do not contribute to these two sectors.
Furthermore, the final normalization term (A.20)
$$N_{20}=ig^2\lambda'\Phi_0^4\delta(x-y)\eqno(5.7)$$ 
becomes now important.
Until now the coupling parameters $b$ (5.1) and $\lambda'$ (5.7) are arbitrary.

Gauge invariance (1.0) at third order can only be violated by local terms
$\sim D\delta (x_1-x_3,x_2-x_3)$ (where $D$ denotes a differential operator).
This can easily be seen by inserting the causal factorization of the
time-ordered products and using gauge invariance at lower orders (cf. (4.3)
in \cite{1}). We only consider the tree diagrams
$$d_QT_3\vert_{\rm tree}=i\sum_{k=1,2,3}\d_\nu^k T_{3/k}^\nu \vert_{\rm tree}
\eqno(5.8)$$
and adopt the notations and terminology from the corresponding calculation 
at second order (sect. 4 of \cite{1}).\footnote{Note that e.g. a term
$\sim\delta(x_1-x_2)\d_\mu\d_\nu D_F(x_2-x_3)$ ($\mu$ and $\nu$ not
contracted) is non-local.} There are no 
local terms in $T_3\vert_{\rm tree}$ and $T_{3/k}\vert_{\rm tree}$, because 
a term $\sim\delta (x_1-x_3,x_2-x_3):B_1B_2B_3B_4B_5:$ would violate 
renormalizability by power counting (terms with derivatives on $\delta 
(x_1-x_3,x_2-x_3)$ are even worse); especially 
there is no freedom of normalization (cf. (4.2) in \cite{1}). Hence the only
local terms in (5.8) are the anomalies which are the terms $\sim\delta 
(x_1-x_3,x_2-x_3)$ in $\sum_{k=1,2,3}\d_\nu^k T_{3/k}^\nu \vert_{\rm tree}
(x_1,x_2,x_3)$. (Derivatives of the $\delta$-distribution do not appear
as can be seen by power counting.) Up to permutation of the vertices there
is only one possibility to generate an anomaly: due to causal factorization,
e.g. $T_{3/1}^\mu (x_1,x_2,x_3)=T_{1/1}^\mu (x_1) T_2(x_2,x_3)$ for
$x_1\not\in(\{x_2,x_3\}+\bar V^-)$, the normalization term $N_{(2)}(x_2,x_3)
\sim\delta (x_2-x_3)$ of $T_2\vert_{\rm tree}(x_2,x_3)$ is contained in
$T_{3/1}^\mu (x_1,x_2,x_3)$. Hence, if $\d^\mu\Phi_0$ in $D_1$, or 
$\d^\mu\Phi$ in $D_2$ (5.2), is contracted with a field operator in 
$N_{(2)}(x_2,x_3)$ we obtain a term
$$T_{3/1}\vert_{\rm tree}^\mu (x_1,x_2,x_3)=\sim\d^\mu D_F(x_1-x_2)\delta
(x_2-x_3):B_1(x_1)B_2(x_1)B_3(x_2)B_4(x_3)B_5(x_3):+...\>.$$
Taking now the divergence with respect to the $Q$-vertex $x_1$, an anomaly 
appears due to \break $\d_\mu\d^\mu D_F(x_1-x_2)=-m^2D_F(x_1-x_2)+\delta
(x_1-x_2)$. Considering the external field operators $u\Phi\Phi_0^3$
there come anomalies from the combinations of $D_1$ with $N_{20}$ and of
$D_2$ with $N_2$. Gauge invariance requires the cancellation of these
anomalies 
$$(D_1,N_{20})_{\rm loc}+(D_2,N_2)_{\rm loc}=0,$$ 
which gives the condition 
$$-i{g\over 2}\>4ig^2\lambda'+i{g\over
2}\>2ig^2\Bigl( {m_H^2\over 4m^2}-{3b\over 2m}\Bigl)=0.\eqno(5.9)$$ 
The factors 4 and 2 are due to the number of possible contractions between 
the two members $D$ and $N$. Because $N_{20}$ and $N_2$ depend on the
unknown parameters $\lambda^\prime$ and $b$, this is a first equation to 
determine these parameters.

The anomalies with field operators $u\Phi\Phi_0W_\nu W^\nu$ come from the
terms 
$$(D_1,N_1)_{\rm loc}+(D_2,N_{10})_{\rm loc}=0.$$ 
This gives the condition
$$-i{g\over 2}\>2i{g^2\over 4}+i{g\over 2}\>2{i\over 4}g^2=0,$$ which is
automatically satisfied. Finally, in the sector $u\Phi_0 \Phi^3$ gauge
invariance requires
$$(D_1,N_2)_{\rm loc}+(D_2,N_{11})_{\rm loc}=0$$ 
and this leads to the constraint 
$$-i{g\over 2}\>2ig^2\Bigl({m_H^2\over 4m^2}-{3b\over
2m}\Bigl) +i{g\over 2}\>4(-ig^2){m_H^2\over 16m^2}=0.\eqno(5.10)$$ 
>From (5.9) and (5.10) we find the values of the two coupling parameters 
$$b={m_H^2\over 4m},\quad\lambda'=-{m_H^2\over 16m^2}.\eqno(5.11)$$

The same argument applies to the whole electroweak theory: In the sector
$u_1\Phi_1\Phi_0^3$ we get the equation (5.9) (with $m=m_W$), and in the
sector $u_1\Phi_0\Phi_1^3$ we find the condition (5.10) again. It is easy to
see that there are no additional terms contributing in the bigger theory. The
resulting values $$b={m_H^2\over 4m_W},\quad\lambda'=-{m_H^2\over
16m_W^2}\eqno(5.12)$$ agree with those obtained from the Higgs potential in the
standard theory as shown below. We have verified that, with these parameters,
all other anomalies from third order tree graphs cancel out, without giving
further information.

For comparison with the standard model we collect all self-couplings of scalar
fields. The first order terms are contained in (3.20.10-12) of \cite{1}:
$$V_1(\Phi)=ig{m_H^2\over 4m_W}\Phi_0(\Phi_0^2+\Phi_1^2+\Phi_2^2
+\Phi_3^2).\eqno(5.13)$$ The second order terms are given by the normalization
terms in the appendix of \cite{1} $$V_2(\Phi)=ig^2{m_H^2\over
16m_W^2}\Bigl[-(\Phi_0^2+\Phi_1^2+\Phi_2^2 +\Phi_3^2)^2\Bigl].\eqno(5.14)$$
Remembering that the second order must be multiplied by 1/2, we obtain the
following total scalar potential $$V(\Phi)=V_1+{1\over 2}V_2=-ig^2{m_H^2\over
32m_W^2}\Bigl[(\Phi_0^2+ \Phi_1^2+\Phi_2^2+\Phi_3^2)^2-8{m_W\over
g}\Phi_0(\Phi_0^2+\Phi_1^2 +\Phi_2^2+\Phi_3^2) \Bigl].\eqno(5.15)$$ To compare
this with the Goldstone-Higgs potential of the standard theory we set
$\Phi_1=\Phi_2=\Phi_3=0$ and add a mass term ${1\over 2}m_H^2\Phi_0^2$.  
 Omitting the factor $-i$, the potential is then
equal to $$V(\Phi_0)=\lambda(\Phi_0^4-4a\Phi_0^3+4a^2\Phi_0^2),\eqno(5.16)$$
where $$\lambda={1\over 2}\biggl({gm_H\over 4m_W}\biggl)^2,\quad a={2\over
g}m_W.\eqno(5.17)$$ This is the shifted double-well potential discussed in the
introduction (1.1). Our point is that this structure is not obtained from a
clever choice of a Lagrangean and subsequent symmetry breaking, but comes out as
the necessary consequence of gauge invariance in the massive situation.
\vfill\eject

 \end{document}